\documentclass[osajnl,twocolumn,showpacs,superscriptaddress,floatfix,10pt]{revtex4-1} 
\usepackage{amsmath,amssymb,graphicx}

\def\wisk#1{\ifmmode{#1}\else{$#1$}\fi}

\def\amin   {\wisk{^\prime\ }}

\def\deg    {\wisk{^\circ}}

\begin{document}

\title{Polarization Properties of A Broadband Multi-Moded Concentrator}

\author{Alan Kogut}
\affiliation{NASA Goddard Space Flight Center, Greenbelt, Maryland 20771, USA}

\author{Dale J. Fixsen}
\affiliation{NASA Goddard Space Flight Center, Greenbelt, Maryland 20771, USA}

\author{Robert S. Hill}
\affiliation{NASA Goddard Space Flight Center, Greenbelt, Maryland 20771, USA}


\begin{abstract}
We present the design and performance 
of a non-imaging concentrator
for use in broad-band polarimetry at millimeter 
through submillimeter wavelengths.
A rectangular geometry preserves the input polarization state
as the concentrator
couples $f/2$ incident optics
to a $2\pi$ sr detector.
Measurements of the co-polar and cross-polar beams
in both the few-mode and highly over-moded limits
agree with
a simple model 
based on mode truncation.
The measured co-polar beam pattern 
is nearly independent of frequency
in both linear polarizations.
The cross-polar beam pattern
is dominated by a uniform term 
corresponding to polarization efficiency 94\%.
After correcting for efficiency,
the remaining cross-polar response is -18 dB.
\end{abstract}


\maketitle 

\section{Introduction}
The advent of bolometric detectors
with background-limited sensitivity
has important implications
for astronomical instrumentation
at millimeter and submillimeter wavelengths.
Once the detector phonon noise
falls below the background
from photon statistical fluctuations,
further sensitivity gains
can only be realized by collecting additional photons.
A common implementation
increases the effective detecting area
through an array of individual sensors,
each coupled to the sky
through an optical structure
(feed horn, lenslet, phased antenna array)
that restricts the system response
to a single electromagnetic mode at the sensor.
Single-moded systems achieve
diffraction-limited angular resolution
with well-defined (Gaussian) beam profiles,
but the large number of sensors required
drives system-level complexity and cost.

An alternative design uses a much smaller number
of detectors capable of sensing multiple modes of the incident field.
A multi-moded system
uses non-imaging optics
(a ``light bucket'' or similar feed structure)
to fix the beam size independent of wavelength.
Since the Lagrange invariant is conserved,
in order for the beam size to remain constant
the number of modes must scale with wavelength,
$N_{\rm mode} = A \Omega / \lambda^2$,
where
$A$ is the detector area,
$\Omega$ is the solid angle,
and
$\lambda$ is the observing wavelength.
The different electromagnetic modes
form an orthogonal basis set
and add incoherently at the detector.
The number of detectors required to reach a given sensitivity
thus scales with etendu
as $(A \Omega)^{-1}$,
inversely proportional to the number of modes.

Multi-moded systems can be particularly useful
for observations of the linear polarization
of the cosmic microwave background (CMB),
where the signals of interest
lie well below the photon background noise.
Polarimetry adds an additional level of complexity
to the optical system.
Multi-moded systems
typically require a concentrator
or similar feed structure
to couple light from the sky
to the detector
while conserving etendu.
Polarization sensitivity can be achieved using a
detector sensitive to both polarizations, 
with a polarization diplexer 
({\it e.g.} a wire grid analyzer) upstream of the detector.
Since the polarization separation occurs before 
light enters the concentrator,
the polarization properties of the concentrator do not matter.
Alternatively, the detector itself may be sensitive to
a single linear polarization.
This avoids the need for a separate polarization diplexer,
but requires 
the concentrator to preserve the
incident polarization state
without cross-polar mixing
(see, {\it e.g.}, the discussion in \cite{kusaka/etal:2014}).
Instruments in this latter class
include the planned
PIXIE \citep{kogut/etal:2011},
MUSE \citep{kusaka/etal:2012}
and LSPE \citep{aiola/etal:2012}
missions.

The design of non-imaging concentrators
has been well studied
for unpolarized observations
\citep{winston:1970,
winston/welford:1979,
minano:1985,
ogallagher/etal:1987,
ning/etal:1987,
garcia-botella/etal:2009}.
Three-dimensional concentrators 
formed as a surface of revolution
can approach the theoretical limit on concentration,
but the resulting azimuthal symmetry
induces an undesirable cross-polar response.
In this paper,
we describe the design and performance
of a non-imaging concentrator
using a rectangular geometry
suitable for CMB polarimetry.

\begin{figure*}[t]
\centerline{
\includegraphics[height=7cm]{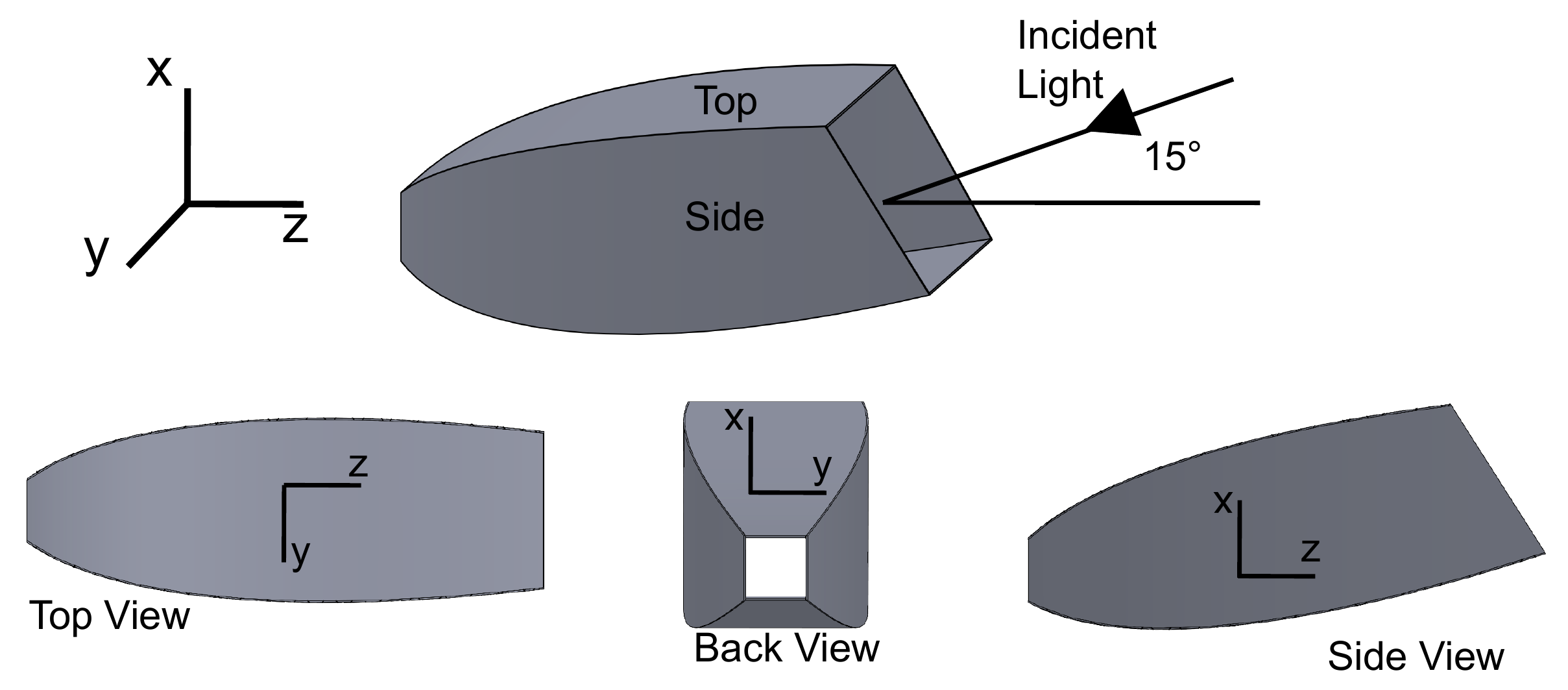}
}
\caption[Schematic of PIXIE concentrator]
{
Schematic of the off-axis non-imaging rectangular concentrator. 
The square entrance and exit apertures
preserve incident polarization.
Each of the 4 walls is an elliptical section
to transform the $f/2$ incident beam
to the $2\pi$ sr solid angle at the detector
while preserving etendu.
}
\label{feed_schematic}
\end{figure*}

\section{Multi-moded rectangular concentrator}

Figure \ref{feed_schematic} illustrates the concentrator,
derived from the requirements of the PIXIE instrument.
We define a right-handed coordinate system $[x, y, z]$
with the detector centered in the $xy$ plane 
and the concentrator extending in the $+z$ direction.
Light from the $f/2$ beam-forming optics 
is incident on the concentrator
at a 15\deg ~angle from the $\hat{z}$ axis.
Within these constraints,
we use a rectangular geometry for the concentrator
to minimize cross-polar mixing of the incident light.
Each of the 4 walls of the concentrator
is an elliptical section,
optimized from an iterative ray-tracing algorithm
to maximize the total throughput
in the geometric optics limit.
We define the ellipse in each plane
as a quadratic equation
\begin{widetext}
\begin{eqnarray}
M_1(x - c_1)^2 + M_2(z - c_2)^2 + 2M_3(x - c_1)(z - c_2) &=& 1 ~~ ({\rm top~face}) 	\nonumber	\\
M_1(x - c_1)^2 + M_2(z - c_2)^2 + 2M_3(x - c_1)(z - c_2) &=& 1 ~~ ({\rm bottom~face})  	\nonumber 	\\
M_1(y - c_1)^2 + M_2(z - c_2)^2 + 2M_3(y - c_1)(z - c_2) &=& 1 ~~ ({\rm side~faces}) 
\label{ellipse_def}
\end{eqnarray}
\end{widetext}
for the top, bottom, and side faces, respectively.
Table 1 lists the coefficient values $M$
and $c$.
The two side panels 
($+\hat{y}$ and $-\hat{y}$)
are identical,
while the top and bottom panels
($+\hat{x}$ and $-\hat{x}$)
have different elliptical sections
to accommodate the off-axis design.
The projected entrance aperture
thus appears approximately square.

\begin{table*}[t]
\caption{Concentrator Elliptical Figure}
\begin{center}
\label{ellipse_table}
\begin{tabular}{l c c c c c}
\hline
Face	& $c_1$ & $c_2$ & $M_{1}$  & $M_{2}$  & $M_{3}$ \\
\hline
Top	& -2.6043E+01 &  4.9987E+02 &  1.3455E-03  & 1.0025E-04  &  1.7423E-05	\\
Bottom  &  3.5040E+02 &  6.0139E+02 &  5.3181E-04  & 1.9821E-04  & -2.6392E-04	\\
Left	& -1.7033E+02 &  5.6460E+02 &  8.7479E-04  & 1.3043E-04  & -2.1488E-04	\\
Right	&  1.7033E+02 &  5.6460E+02 &  8.7479E-04  & 1.3043E-04  &  2.1488E-04	\\
\hline
\multicolumn{6}{c}{Note: Units are in mm.  Values describe the larger
scaled concentrator ($\S$3).}
\end{tabular}
\end{center}
\end{table*}

The exit aperture of the concentrator
illuminates a pair of detectors,
each sensitive to a single linear polarization.
Each detector covers the full aperture;
they are mounted one behind the other
so that both are fully illuminated.
The two detectors are rotated by 90\deg
~such that one detector is sensitive to polarization 
in the $\hat{x}$ direction
and the other is sensitive to polarization 
in the $\hat{y}$ direction.
Each detector has etendu 
$A \Omega = 4$ cm$^2$ sr.
We model the co-polar and cross-polar response 
of the concentrator using
a time-reversed Monte Carlo approach.
For each linear polarization,
we generate a set of 63 million outgoing rays,
each originating from a random position on the detector.
The ray angular distribution 
is uniform in azimuth,
but takes into account
the additional $cos^2(\theta)$ angular response
of a thin resistive detector
at the concentrator exit aperture\cite{kusaka/etal:2014}.
The distribution in polarization is uniform.
A ray-tracing algorithm
follows each ray
through multiple bounces within the concentrator
(assumed to be perfectly reflective)
to determine the orientation
and polarization angle
after leaving the concentrator.

The PIXIE concentrator does not view the sky directly,
instead coupling to the beam-forming optics
through an $f/2$ coupling mirror.
We extend the ray-tracing algorithm
to follow each ray 
out of the concentrator
and and through one reflection from the coupling mirror.
The resulting set of rays
approximates the beam pattern
of the concentrator/coupling mirror combination,
which can then be used as input
through the remaining optical system.
The co-polar and cross-polar response 
of any downstream beam-forming optics
can readily be calculated,
and and not treated here.

\begin{figure}[b]
\centerline{
\includegraphics[width=0.9\columnwidth]{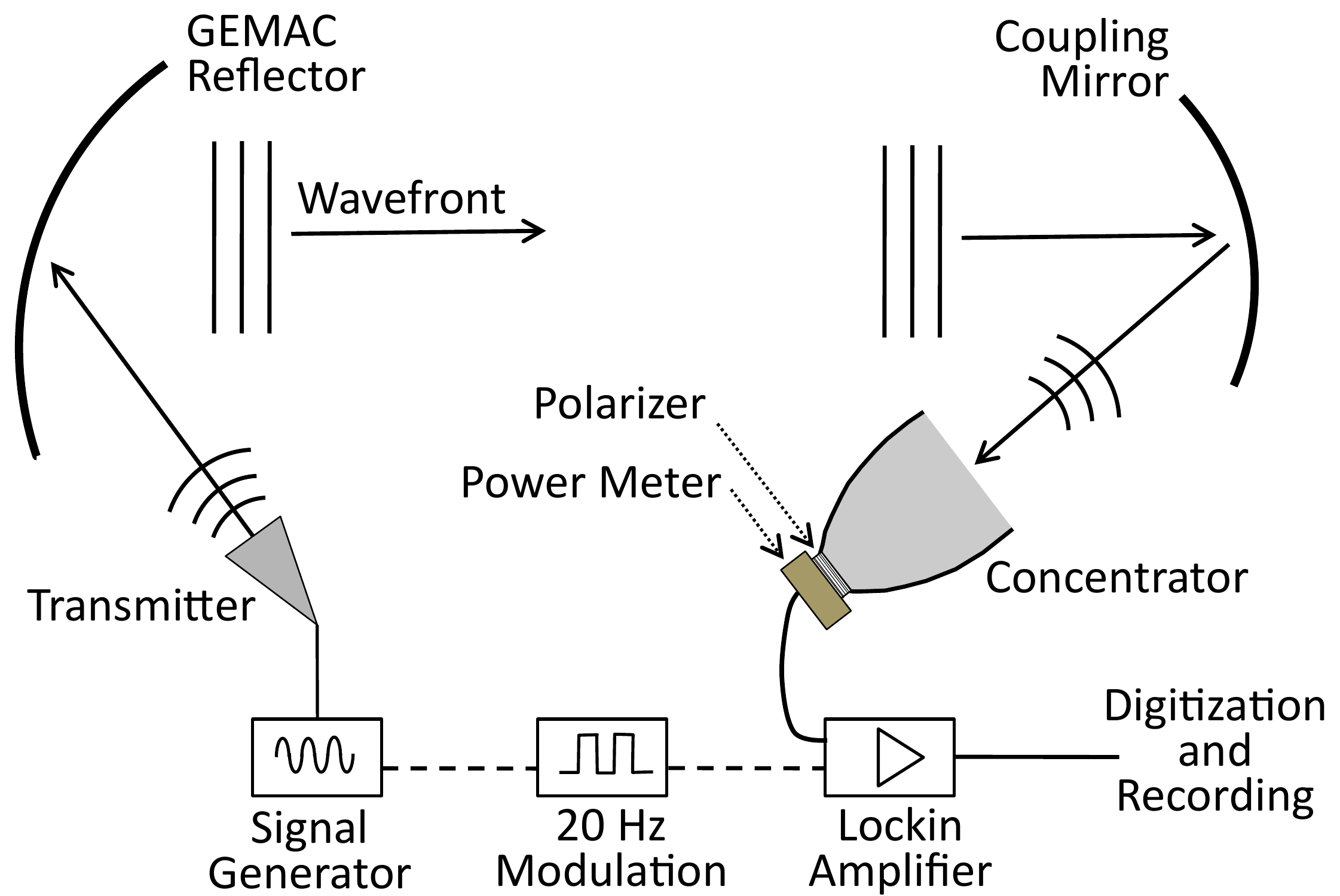}
}
\caption[Block diagram]
{
Block diagram for beam pattern measurement.
}
\label{gemac_fig}
\end{figure}

Observations of the cosmic microwave background
can span more than a decade in frequency,
from 30 GHz to 600 GHz.
The corresponding number of modes 
within the concentrator
ranges from
$N_{\rm mode} = 4$
at the lowest frequency
to $N_{\rm mode} = 1600$ at the highest frequency.
The concentrator
thus samples the transition
from a few-mode system 
to a highly over-moded system.
Several authors have discussed
methods to estimate the beam pattern
for non-imaging concentrators 
in the few-mode limit
\citep{murphy/padman:1991,
gleeson/etal:2002,
thomas/withington:2013}.
Typically these methods require
explicit mode matching.
Although mode matching successfully recovers
the full phase and amplitude information
needed to generate far-field beam patterns,
this method is computationally expensive
and is difficult to apply
for off-axis geometries
where a full 3-dimensional treatment is required.
To compare the modeled beam patterns
to measurement,
we employ a simpler approximation.
We use the beam pattern from the
ray-tracing algorithm to model the beam
in the geometric optics limit.
We then mathematically expand the beam 
using an orthogonal set of basis functions,
and recompute the beam
using a truncated subset of these functions
to account for mode loss at low frequencies.
The rectangular symmetry of the concentrator
is well matched 
to the orthogonal basis functions
of a two-dimensional Fourier transform.
We thus bin the co-polar or cross-polar beam
derived from the geometric optics model
on a $6\amin \times 6\amin$ grid,
and Fourier transform the binned beam map
to derive a set of complex Fourier coefficients.
We sort the coefficients by angular frequency
and retain only the lowest $N_{\rm mode}$ values,
setting the remaining modes to zero.
We then Fourier transform back to real space
to generate a model with the desired number of modes.
Finally, we convolve the mode-truncated pattern
with the Airy pattern for the appropriate observing frequency
to approximate the effects of diffraction at the concentrator walls.

\begin{figure}[b]
\centerline{
\includegraphics[width=0.9\columnwidth]{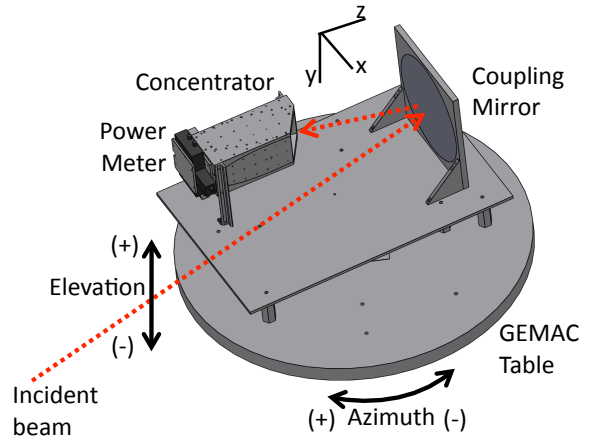}
}
\caption[Table vs horn coordinates]
{
Schematic showing the concentrator and coupling mirror
mounted for measurement in the GEMAC azimuth/elevation table,
showing the relation between
the table and concentrator coordinate systems.}
\label{coord_figure}
\end{figure}

\begin{figure*}[t]
\centerline{
\includegraphics[width=1.7\columnwidth]{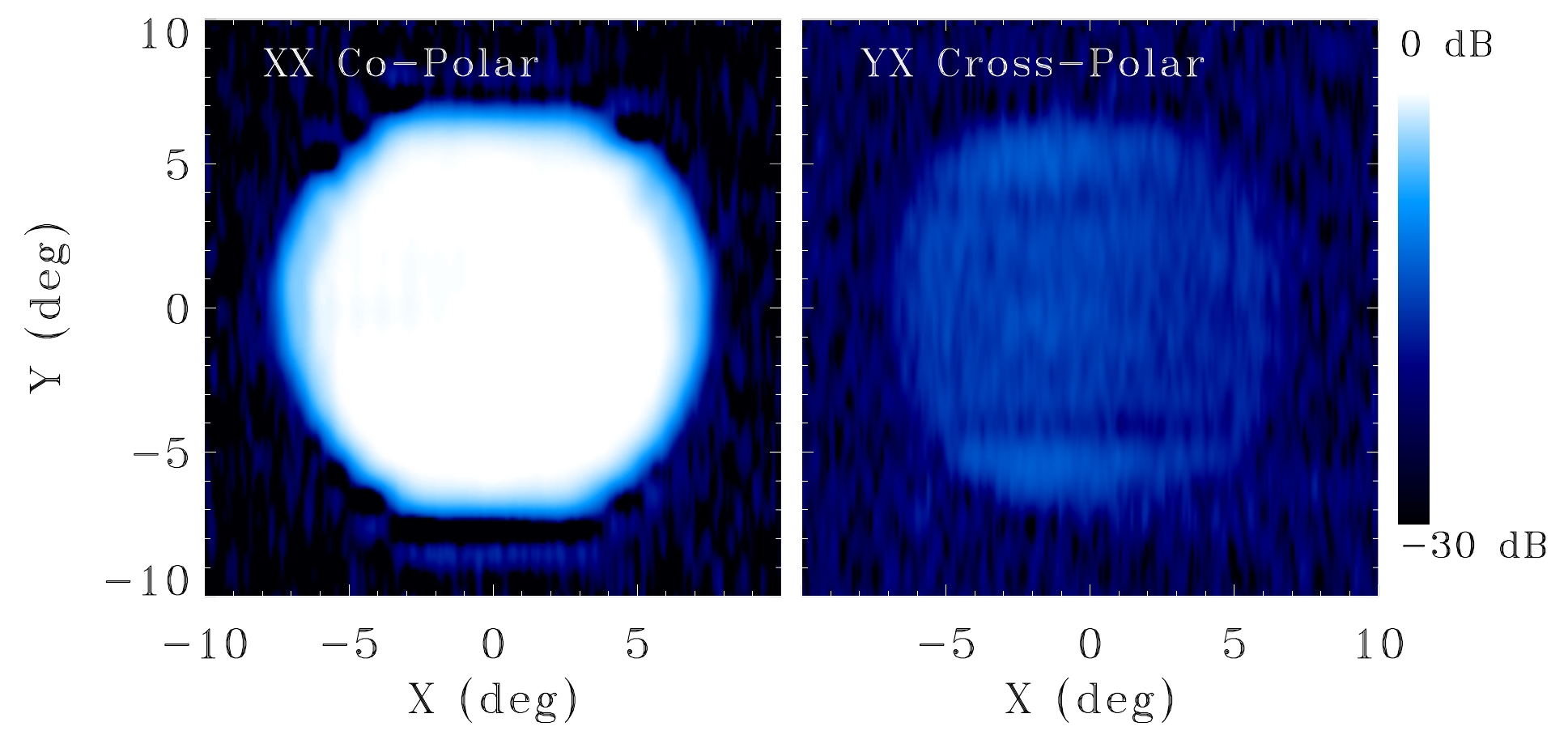}
}
\caption[Beam Maps at 90 GHz]
{
Co-polar and cross-polar beam maps 
of the scaled concentrator / coupling mirror system.
The measurement of the scaled concentrator at 90 GHz 
corresponds to 270 GHz for PIXIE,
well within the geometric optics limit.
The transmitter is polarized in the $\hat{x}$ direction
for the copolar map and the  $\hat{y}$ direction
for the cross-polar map.
The co-polar beam is well described 
by a tophat with diameter 14\deg.
The cross-polar beam is dominated
by a similar tophat at -12 dB
(corresponding to polarization efficiency 0.94),
with 
smaller-scale structure at the -18 dB level.
}
\label{gemac_greyscale}
\end{figure*}

\section{Beam pattern measurements}

We measured the co-polar and cross-polar beam patterns 
of the rectangular concentrator
in the Goddard Electromagnetic Anechoic Chamber (GEMAC).
Figure \ref{gemac_fig} shows the experimental setup.
A transmitter within the GEMAC 
uses a standard rectangular gain feed
to launch a single linear polarization
toward a shaped reflector.
The reflector transforms the outgoing signal
to a plane wave,
effectively placing the transmitter at infinity.
We include a coupling mirror
in the GEMAC setup
to convert the transmitted plane wave
to the $f/2$ beam generated by the PIXIE coupling mirror,
and measure the copolar and cross-polar beams
from the concentrator/mirror system.
The GEMAC coupling mirror differs
from the PIXIE mirror
in that the PIXIE mirror 
uses an elliptical surface 
to focus on the next mirror in the instrument
while the GEMAC coupling mirror
uses a parabolic surface
to focus on the transmitter at infinity.
Since the GEMAC setup excludes the remainder of the beam-forming optics,
the measured beam patterns represent
only the concentrator/coupling mirror combination
and are not a measure of the final instrument beams on the sky.

The concentrator/mirror system is mounted
on a table 
and can be moved in azimuth and elevation.
Figure \ref{coord_figure} show the relative
orientation of the table and concentrator coordinate systems.
Viewed from the detector in the concentrator $xy$ plane,
moving the table towards positive azimuth
moves a source at infinity
toward 
positive $\hat{x}$
(as viewed through the coupling mirror).
Similarly, moving the table towards positive elevation
moves a source at infinity
toward positive $\hat{y}$.
To suppress reflected signals,
all surfaces except the coupling mirror
and concentrator aperture
were covered with a single layer
of Eccosorb AN72 microwave absorber.
The absorber at the edges of the coupling mirror
acts as a beam stop,
truncating the corners
to produce the desired tophat geometry.

We modulate the transmitted power 
by electronically switching the transmitted power
with a square wave at 20 Hz,
and use a Thomas Keating THz absolute power meter
\cite{tk_meter}
at the exit aperture of the concentrator
to detect and synchronously demodulate the signal.
The power meter is insensitive to polarization;
a free-standing wire grid polarizer
mounted between the power meter and the concentrator exit aperture
provides polarization sensitivity.
The polarizer is flush to the concentrator exit aperture
while the power meter is parallel to the exit aperture
but 3 mm further back within a reflective integrating cavity.
The wire grid polarizer consists of 
copper-clad tungsten wires
40 microns in diameter
(36 $\mu$m tungsten and 1.3--2.5 $\mu$m copper)
on a 118 $\mu$m pitch.
With the grid wires oriented parallel to the elevation direction,
the $\hat{y}$ polarization is reflected and
the power meter is sensitive to $\hat{x}$ polarization.
The grid can be rotated by 90\deg
~so that the wires are parallel to the azimuth direction.
In this orientation,
the $\hat{x}$ polarization is reflected and
the power meter is sensitive to $\hat{y}$ polarization.
The measured polarization isolation
is better than -27 dB
at frequencies 10 GHz to 300 GHz.

\begin{table}[b]
\label{power_table}
\begin{center}
\caption{Measured Map Power}
\begin{tabular}{l c c c c }
\hline
Frequency	& Max Co-Pol & Max X-Pol & Noise Co-Pol & Noise X-Pol \\
(GHz)		& ($\mu$W)   & ($\mu$W)  & ($\mu$W)     & ($\mu$W) \\
\hline
11		& 1560	     & 91	 & 1.4		& 0.9	\\
29		& 700	     & 73	 & 0.7	 	& 0.5	\\
90		& 210	     & 24	 & 0.4		& 0.5	\\	
\hline
\end{tabular}
\end{center}
\end{table}

With the transmitter fixed to broadcast $\hat{x}$ polarization,
we orient the polarizer grid parallel to the elevation direction
to measure the XX co-polar response.
We fix the elevation angle at $-10\deg$
and continuously sweep in azimuth
from $-12\deg$
~to $+12\deg$,
recording data from
$-10\deg$
~to $+10\deg$
to avoid possible startup transients in the azimuth drive.
We then increment the elevation by 1\deg
~and repeat
to map the concentrator/mirror response over a 
$20\deg \times 20\deg$ field.
We then rotate the transmitter
to the $\hat{y}$ polarization
while leaving the polarizer grid parallel to the elevation direction
to measure the YX cross-polar response.
Finally, we rotate the wire grid polarizer by 90\deg
~and repeat
to measure the YY co-polar and XY cross-polar response.
Since the off-axis concentrator is slightly different in the 
$\hat{x}$ and $\hat{y}$ directions,
we expect slightly different responses
for the 4 beam patterns.

We wish to measure the beam patterns
at frequencies representative of both
the few-mode and highly-overmoded limits.
The limited broadcast power available at frequencies above 100 GHz
makes direct beam mapping at these frequencies impractical.
We thus increase the dimensions of the nominal PIXIE concentrator
by a factor of three,
and measure the larger (scaled) concentrator
at lower frequencies
where higher broadcast power can be achieved.
We measured all four beam patterns (XX, YX, YY, and XY)
at each of frequencies 11 GHz, 29 GHz, and 90 GHz,
corresponding to CMB frequencies 33, 87, and 270 GHz.
Table 2 summarizes the measurements.
The maximum co-polar power measured at each frequency
ranged from 
1500 $\mu$W at 11 GHz
to 
210 $\mu$W at 90 GHz
with rms noise
1.4 $\mu$W to 0.4 $\mu$W
per 0.1\deg~pixel.
The noise floor is thus at approximately -30 dB
at each frequency.

\begin{figure*}[h]
\begin{center}
\begin{tabular}{c}
\includegraphics[width=2\columnwidth]{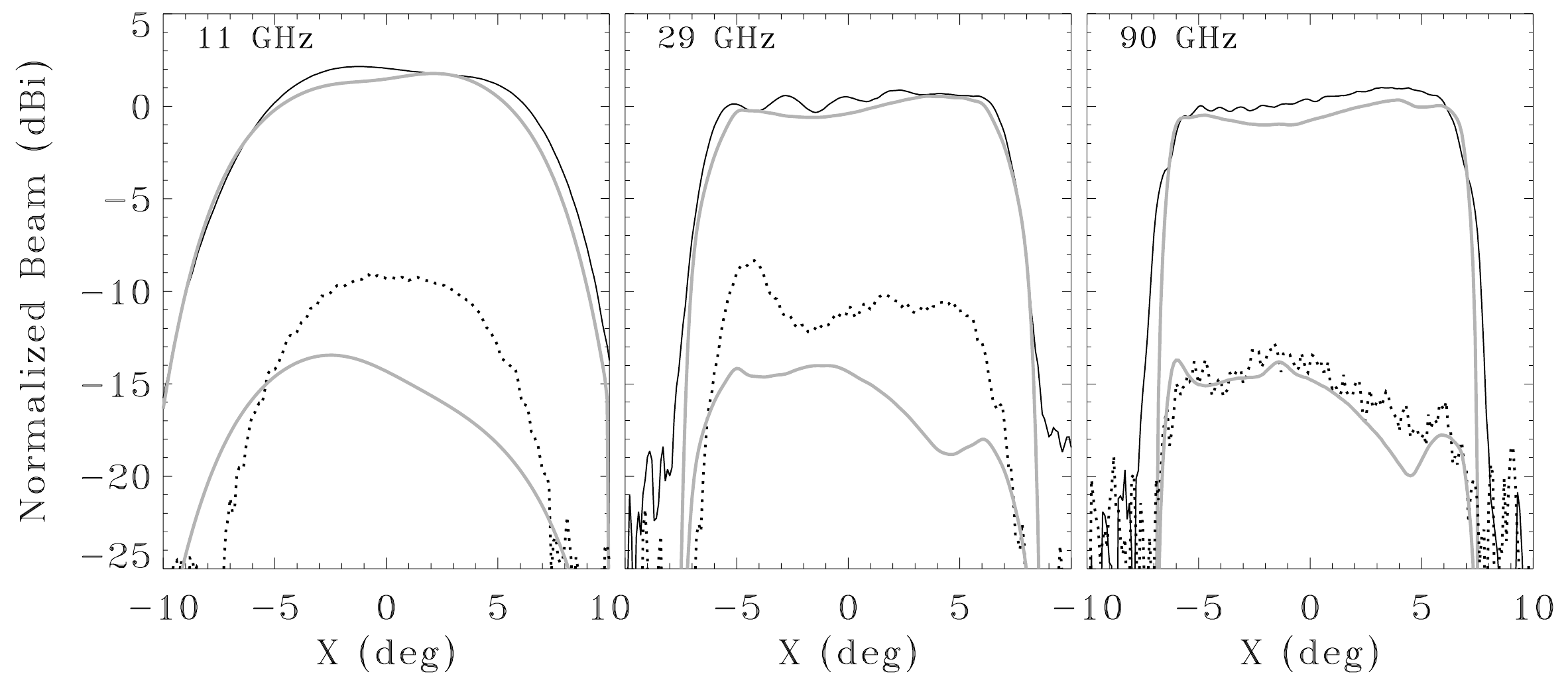}
\end{tabular}
\end{center}
\caption[Measured Azimuth Beams vs Model Predictions]
{Measured beam patterns for 
the scaled concentrator/coupling mirror system
compared to model predictions in the $\hat{x}$ direction.
The solid black line shows the 
co-polar (XX) beam 
while
the dotted line shows the cross-polar (YX) beam.
Data are taken with $\hat{y}$ fixed at 0\deg.
Grey lines show the model predictions
after accounting for diffraction and mode loss.
Measurements at 11, 29, and 90 GHz
correspond to 5, 34, and 324 modes respectively,
spanning the transition
from a few-mode system to a highly over-moded system.
}
\label{slice_fig_az}
\end{figure*}

\begin{figure*}[b]
\begin{center}
\begin{tabular}{c}
\includegraphics[width=2\columnwidth]{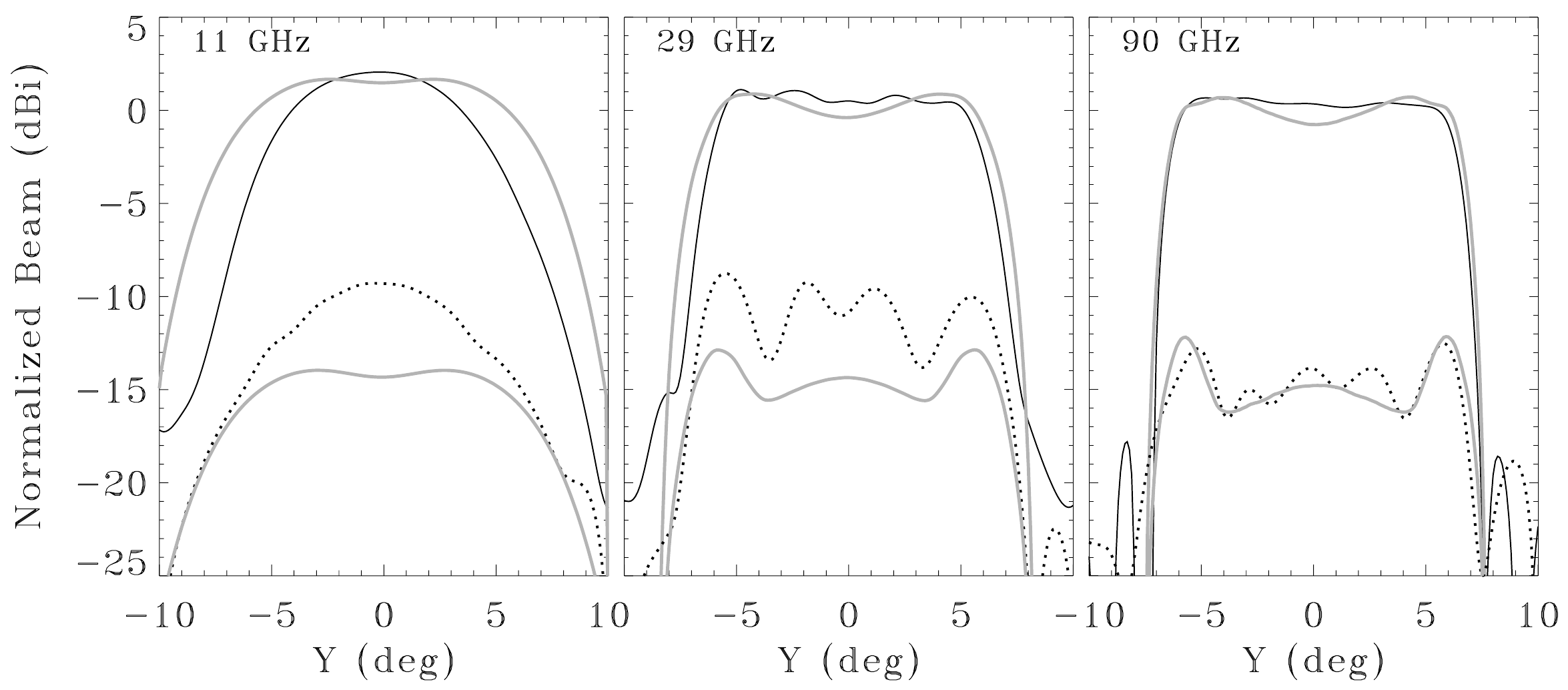}
\end{tabular}
\end{center}
\caption[Measured Elevation Beams vs Model Predictions]
{Measured and model beam patterns
for 
the copolar (XX) and cross-polar (YX) beams
in the $\hat{y}$ direction.
Data are taken with $\hat{x}$ fixed at 0\deg.
Model and measured data are as in Fig \ref{slice_fig_az}.
}
\label{slice_fig_el}
\end{figure*}

At each frequency,
we define the co-polar and cross-polar beam patterns as
\begin{eqnarray}
G_c(x,y) &=& \frac{ P_c(x,y) \Omega}{\int P_c(x,y) \, d\Omega}	\\
G_x(x,y) &=& \frac{ P_x(x,y) \Omega}{\int P_c(x,y) \, d\Omega}
\label{beam_def}
\end{eqnarray}
where
$P_c(x,y)$ is the co-polar power 
measured at each point $(x,y)$,
$P_x(x,y)$ is the measured cross-polar power,
and
\begin{equation}
\Omega = \frac{ \left( \int P_c(x,y) \, d\Omega \right)^2 }
	      { \int P^2_c(x,y) \, d\Omega }
\label{omega_def}
\end{equation}
is the co-polar beam solid angle.
With this choice of normalization
the beam solid angle is simply related
to the beam pattern,
\begin{equation}
\Omega = \int G_c(x,y) \, d\Omega \, ,
\label{omega_eq}
\end{equation}
but with a 
peak amplitude greater than unity,
$G_{\rm max} > 1$
\cite{kraus:1986}.

\begin{table}[b]
\label{width_table}
\begin{center}
\caption{Beam Width}
\begin{tabular}{c c c c c}
\hline
Frequency	& Co-Pol & Solid & FWHM		& Equivalent 	\\
(GHz)		& Map    & Angle & (deg)	& Tophat 	\\
		&	 & (sr)  & 		& (deg)		\\
\hline
11   &   XX  &   0.0502 &  12.5  & 14.5  \\
11   &   YY  &   0.0515 &  13.4  & 14.7  \\
29   &   XX  &   0.0483 &  12.6  & 14.2  \\
29   &   YY  &   0.0477 &  12.4  & 14.1  \\
90   &   XX  &   0.0470 &  12.7  & 14.0  \\
90   &   YY  &   0.0468 &  12.7  & 14.0  \\
\hline
\end{tabular}
\end{center}
\end{table}

Figure \ref{gemac_greyscale} shows the 
co-polar (XX) and cross-polar (XY) beam pattern
measured at 90 GHz in the GEMAC,
corresponding to 270 GHz for PIXIE.
The co-polar beam 
is described by a tophat
with diameter 14\deg.
To lowest order, 
cross-polar beam resembles the co-polar beam
but with amplitude reduced by 12 dB.
Additional higher-order structure
in the cross-polar beam
is also visible,
reflecting the fourfold symmetry of the concentrator walls.

Figure \ref{slice_fig_az}
compares the measured beam maps
to the model.
Each panel shows 
a one-dimensional slice through the beams
along the $\hat{x}$ (azimuth) axis
at $\hat{y} = 0$.
We show the co-polar XX beam pattern
(transmitter broadcasting in $\hat{x}$ 
and concentrator receiving in $\hat{x}$)
as well as the cross-polar YX pattern
(transmitter broadcasting in $\hat{y}$ 
and concentrator receiving in $\hat{x}$).
The co-polar beams agree well with the model
and show the expected broadening
from mode loss
at low frequencies.
The cross-polar response 
is reduced by $\sim$12 dB from the co-polar response
and is dominated 
by the same tophat structure as the co-polar beam.
We observe similar results 
for slices in elevation ($\hat{x} = 0$)
(Figure \ref{slice_fig_el}).
Note that the right-most panels
in Figures \ref{slice_fig_az} and \ref{slice_fig_el}
correspond to 
horizontal or vertical slices
through the 90 GHz beam
shown in Figure \ref{gemac_greyscale}.
The measured beam patterns
show the qualtitative features
predicted by the mode-truncated model.
Differences between the measured and modeled beam patterns
are dominated by an amplitude scale factor
which increases at lower frequency.
Reflections from the mounting plate
beneath the concentrator/mirror system
(Figure \ref{coord_figure})
could account for such an effect.

\begin{figure}[t]
\begin{center}
\begin{tabular}{c}
\includegraphics[width=1.0\columnwidth]{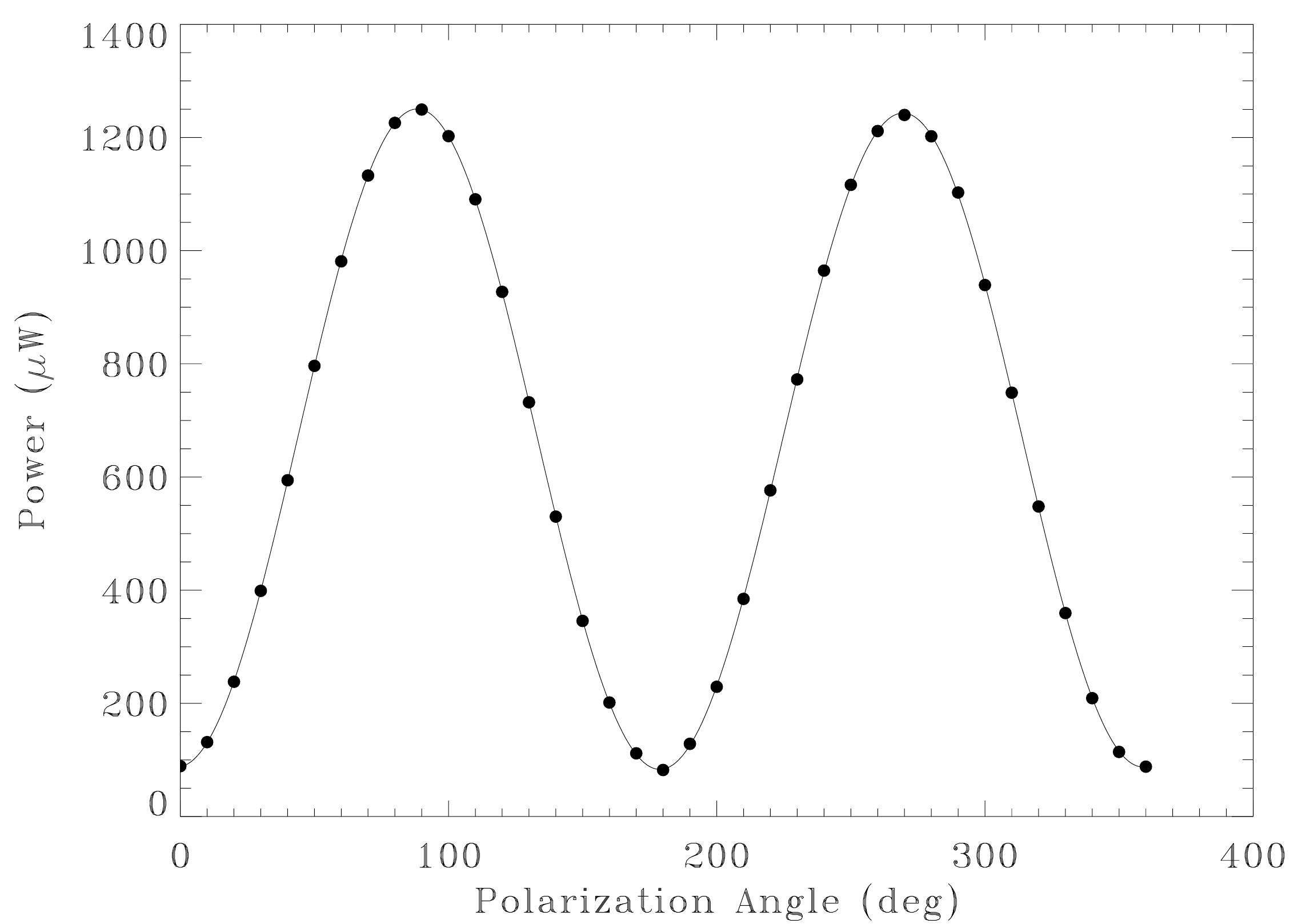}
\end{tabular}
\end{center}
\caption[Polarization Efficiency]
{On-axis response vs polarization angle measured at 11 GHz.
}
\label{pol_eff}
\end{figure}

Diffraction and mode truncation
combine to apodize the beam edges.
The full width at half maximum (FWHM)
of the measured beam is 12.7\deg.
Power outside the half-power point
broadens the beam so that the FWHM
does not fully describe the co-polar beam.
Table 3 shows the FWHM
for each co-polar beam,
along with the diameter of an ideal circular tophat
with the same solid angle as the measured beam.
Although the FWHM is nearly independent of frequency,
the equivalent tophat diameter increases slightly
at the lowest frequency,
showing the effect of mode loss and diffraction.

\subsection{Polarization Efficiency and Cross-Polar Residuals}

A cross-polar response
proportional to the co-polar beam
is equivalent to a loss in polarization efficiency,
but does not 
generate systematic error when mapping the polarized sky.
Figure \ref{pol_eff} illustrates this point.
With the concentrator held stationary,
we record the received power
at 11 GHz
while the transmitter rotates in polarization angle $\theta$.
The measured power shows the expected 
$\sin(2 \theta)$ modulation
with a polarization efficiency 94\%
corresponding to an on-axis cross-polar response of -12 dB.

\begin{table}[b]
\label{pol_table}
\begin{center}
\caption{Polarization Properties}
\begin{tabular}{c c c c c}
\hline
Frequency	& Co-Pol & Cross-Pol & Polarization & Cross-Pol 	\\
(GHz)		& Map    &  Map      & Efficiency   & (dB) 		\\
\hline
11   &   XX  &    YX   &  0.94   &   -19 \\
11   &   YY  &    XY   &  0.95   &   -20 \\
29   &   XX  &    YX   &  0.93   &   -17 \\
29   &   YY  &    XY   &  0.93   &   -15 \\
90   &   XX  &    YX   &  0.97   &   -20 \\
90   &   YY  &    XY   &  0.94   &   -17 \\
\hline
\end{tabular}
\end{center}
\end{table}

\begin{figure}[t]
\begin{center}
\begin{tabular}{c}
\includegraphics[width=1.0\columnwidth]{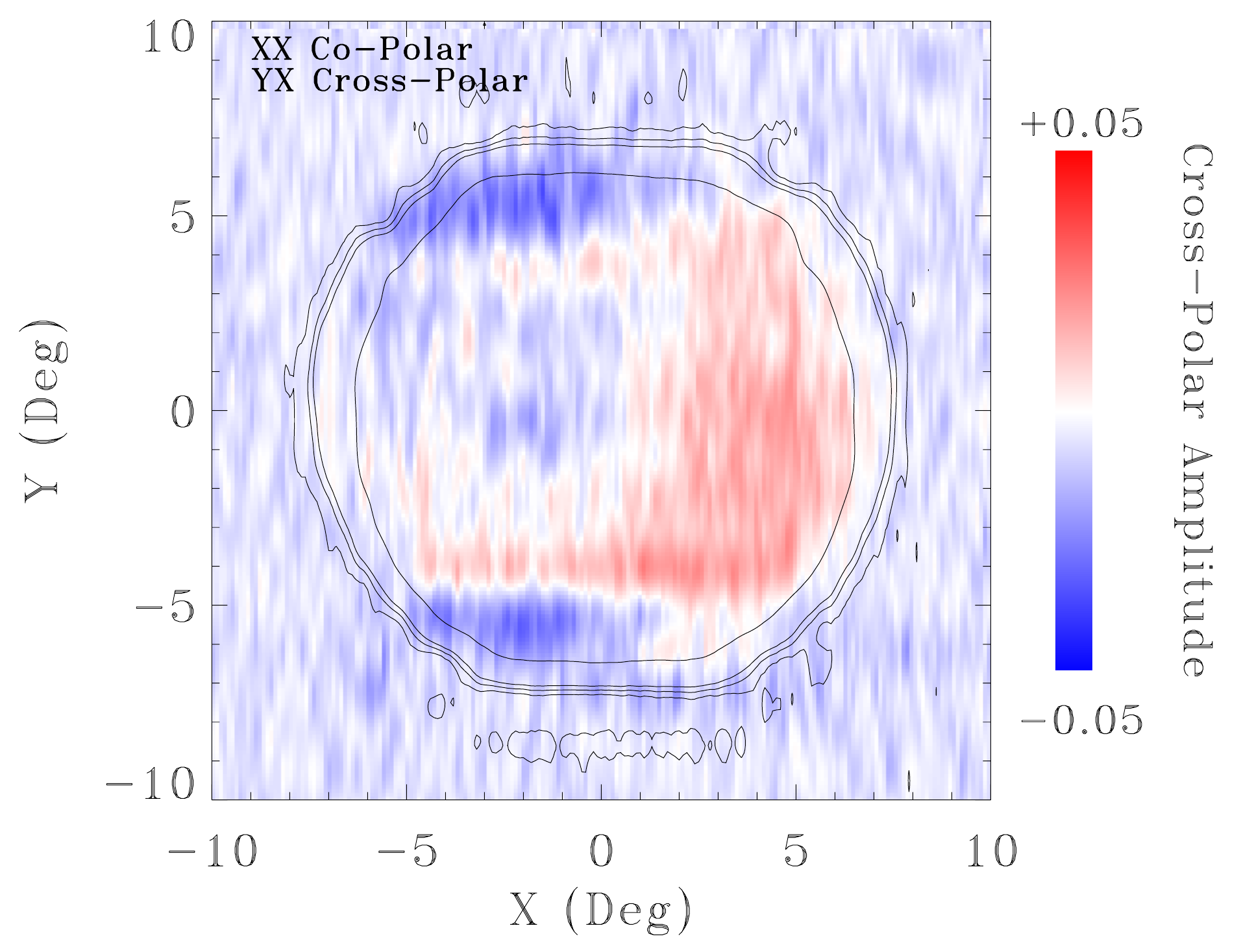}
\end{tabular}
\end{center}
\caption[CO-Polar beam and Cross-polar residual]
{Co-polar beam and cross-polar residual
measured at 90 GHz.
Contours show the XX co-polar beam
on a linear scale at amplitudes 0.01, 0.05, 0.1, 0.5,  and 1.
The YX cross-polar residuals
are shown in color,
with red positive and blue negative.
}
\label{resid_fig}
\end{figure}

We estimate the non-uniform cross-polar response
by fitting each cross-polar beam map
to the corresponding co-polar map,
\begin{equation}
\alpha = \frac{ \int G_x G_c \, d \Omega}{ \int G_c^2 \, d \Omega}
\label{fit_eq}
\end{equation}
and computing the rms amplitude of the residual
\begin{equation}
R_x(x,y) = G_x(x,y) - \alpha G_c(x,y) ~.
\label{resid_eq}
\end{equation}
Figure \ref{resid_fig} compares the
residual cross-polar response
to the co-polar beam
measured at 90 GHz.
The beams are shown on a linear scale
to highlight the cross-polar structure.
The residual cross-polar response
shows the effects of the square aperture and walls.
The asymmetry in the $\hat{x}$ direction
results from the corresponding asymmetry in the concentrator
due to the off-axis design (Figure \ref{feed_schematic}).
Table 4
lists the polarization efficiency
and rms cross-polar residuals
measured for both polarizations
at all three frequencies.
Averaged over the full beam,
the polarization efficiency is 94\%
and the rms cross-polar residual is -18 dB.

\section{Conclusion}
We have designed and tested an off-axis 
rectangular non-imaging concentrator
for use with multi-moded polarization-sensitive optical systems.
The concentrator preserves linear polarization
and conserves the optical etendu
while transforming the beam
from $f/2$ at the entrance aperture
to $2\pi$ sr at the detector ($f/0.5$).
The measured co-polar beam pattern 
is nearly independent of frequency
in both linear polarizations.
The co-polar beam is well described by a tophat
of angular diameter 14\deg
~with minimal internal structure on smaller angular scales.
The cross-polar response is dominated by a uniform tophat,
leading to polarization efficiency 94\%.
After removing the uniform term,
the residual cross-polar response
is typically -18 dB.

Several improvements are possible.
The design process restricted the figure 
for each wall of the concentrator 
to remain an ellipse,
while varying the three foci 
to optimize the total throughput (co-polar and cross-polar).
A more ambitious design
could add additional free parameters to the figure for each wall,
changing the simple ellipse to (e.g.) a low-order polynomial.
Similarly,
the concentrator could be optimized to maximize the co-polar throughput
while simultaneously minimizing the cross-polar response.
We have not yet attempted such improvements.

We measured the co-polar and cross-polar beam patterns
for an optical system consisting of the concentrator
plus a coupling mirror.
The results are thus representative of the beam patterns
expected at output of the PIXIE instrument
and do {\it not} represent the PIXIE beams on the sky.
The full PIXIE optical system
includes additional beam-forming mirrors
to reduce the tophat diameter to the design value 2.6\deg
\cite{kogut/etal:2011}.
The beam-forming optics are optimized 
to further symmetrize the beam patterns.

\acknowledgments
We thank S. Dixon, P. Mirel, and A. Walts
for assistance with beam pattern measurements
and
E. Wollack for fruitful discussion.




\end{document}